\documentclass[
    aps,
    pre,
    reprint,
    superscriptaddress,
    floatfix,
    amsmath, amssymb, amsfonts
]{revtex4-2}
\usepackage{graphicx}
\usepackage[english]{babel}
\usepackage{enumerate}
\usepackage[makeroom]{cancel}
\usepackage{mathtools}
\usepackage[final]{microtype}
\usepackage[caption=false]{subfig}
\usepackage{csquotes}
\usepackage{ulem}
\usepackage[
    colorlinks=true,
    linkcolor=black,
    citecolor=black,
    urlcolor=blue
]{hyperref}
\usepackage[capitalize,nameinlink]{cleveref}

\newcommand{\R}{\mathbb{R}}

\DeclareMathOperator{\const}{const}
\DeclarePairedDelimiter{\mean}{\langle}{\rangle}
\newcommand{\degeneps}{10^{-8}} 

\begin{document}

\title{Collective communication in a transparent world: Phase transitions in a many-body Potts model and a social--quantum correspondence}

\author{Pawat Akara-pipattana}
\affiliation{LPTMS, CNRS -- Universit\'e Paris Saclay 91405 Orsay Cedex, France}

\author{Sergei Nechaev}
\affiliation{LPTMS, CNRS -- Universit\'e Paris Saclay 91405 Orsay Cedex, France}
\affiliation{Beijing Institute of Mathematical Sciences and Applications (BIMSA), Yanqi Lake, Huairou District, Beijing 101408, China}

\author{Bogdan Slavov}
\affiliation{Independent Researcher}

\begin{abstract}
Digitally connected societies approach a ``transparent'' regime where all agents can interact without geographic or social barriers --- a limit realized by complete graph topologies. We construct a mean-field theory of a $q$-state Potts model with many-body interactions on this geometry, modeling agents from $q$ distinct communities.
Analyzing the illustrative case of competing pairwise and three-body couplings, we identify three equilibrium phases in the thermodynamic limit: symmetric (all communities equal), reduced-symmetry ($q-1$ communities surviving), and consensus (one dominant group). For two-community systems, we identify a special coupling regime where the influence of interactions cancels out, yielding purely entropy-driven dynamics --- a statistical physics representation of atomized societies without structured influence. Monte Carlo simulations confirm these regimes and reveal metastable switching dynamics in finite systems. Furthermore, we establish a correspondence ($q \leftrightarrow N$) between this model and a mean-field $SU(N)$ quantum spin system with quadratic and cubic interactions. This ``social--quantum'' correspondence suggests a representation-theoretic description for classifying symmetry-broken macrostates and offers an interpretive link between quantum mean-field phase structure and opinion stratification.
\end{abstract}

\date{\today}

\maketitle

\section{Introduction}
\label{sec:intro}

Collective phenomena emerging from local interactions are central in statistical physics, network science and social dynamics. The Schelling model \cite{schellingDynamicModelsSegregation1971} is a canonical example: agents occupying lattice sites relocate when the local fraction of unlike neighbors exceeds their tolerance. In the original formulation, this repeated relocation process drives the system towards an equilibrium spatially segregated configuration. As the satisfaction (tolerance) threshold is varied, the system undergoes a transition from mixed states to monochromatic clusters with fluctuating boundaries. For mathematical details, generalizations and physical analogies, see, for example, \cite{dallastaStatisticalPhysicsSchelling2008, vinkovicPhysicalAnalogueSchelling2006, hatnaCombiningSegregationIntegration2015, singhCreatingAdaptiveNetwork2006, fagioloSegregationNetworks2007, henryEmergenceSegregationEvolving2011}.

Beyond spatial lattices, our work has explored Schelling dynamics on general networks. Avetisov et al.~\cite{avetisovPhaseTransitionsSocial2018a} studied a variant on graphs, where a society is modeled as a social network with different colors indicating distinct social categories. They showed that competition between two-body interactions among individuals from different groups and three-body interactions within connected triads can reproduce distinct social behaviors. Numerical analysis demonstrated phase transitions between various topologies, including bipartite graphs and clustered graphs linked via \enquote{ambassadors}.

Our work develops these ideas in two directions. First, we formulate the problem as a $q$-state Potts system with many-body couplings on a complete graph. The complete graph topology reflects the limiting case of contemporary \enquote{transparent societies} where digital technologies enable individuals to observe and influence each other across traditional boundaries. Exact description of mean-field Potts models with pairwise interactions have been developed using diffusion equations in \cite{lorenzoniExactAnalysisPhase2019a}. By contrast, our combinatorial approach treats general many-body couplings. Second, our model has some links with the mean-field theory of $p$-star exponential random graph models \cite{biondiniPstarModelsMeanfield2022}, and we discuss this connection in \cref{sec:general-q-and-p-star}. Unlike the original Schelling model where community sizes are fixed, our framework determines equilibrium community sizes $n_1, \ldots, n_q$ by free energy minimization, allowing spontaneous emergence of majority and minority groups.

To solve the model we use the occupation-number variables for finite $n$ and analyze the thermodynamic limit $n \to \infty$ on the complete graph, where description is exact. Phase diagrams are obtained by minimizing the per-site free energy $F_\infty$. Three distinct equilibrium phases emerge: a symmetric phase (all $q$ communities equally populated), a reduced-symmetry phase ($q-1$ equal communities), and a consensus phase (one dominant group with symmetric minorities). These phases are patterns of spontaneous symmetry breaking of the $S_q$ permutation symmetry under color relabeling. We complement analytical results with Monte Carlo simulations that confirm the phases and reveal metastable switching dynamics between degenerate configurations. We also establish a free-energy correspondence with mean-field $SU(N)$ quantum spin systems at $r=3$, which motivates a representation-theoretic classification of macrostates (Young diagrams); sociological readings of this link are discussed in \cref{sec:discussion}.

The paper is organized as follows. In \cref{sec:model} we derive the Hamiltonian and effective potential for the $q$-state Potts model with many-body interactions on a complete graph. In \cref{sec:phase-structure} we analyze the phase structure in the thermodynamic limit for systems with two- and three-body interactions. In \cref{sec:finite-effects} we examine finite-size effects and presents Monte Carlo simulations. In \cref{sec:social-quantum-corr} we establish the social--quantum correspondence. \cref{sec:discussion} provides sociological interpretations, discusses limitations, and suggests open questions for future research.

\section{Model and Methods}
\label{sec:model}

\subsection{Hamiltonian in terms of occupation numbers}
\label{sec:model-hamiltonian-occup}

We begin by deriving an exact representation of the Hamiltonian of $n$ nodes with $q \geq 2$ distinct community types in terms of occupation numbers. The configuration space is thereby reduced from $q^n$ microstates to $O(n^{q-1})$ occupation-number partitions.

We consider a $q$-state Potts model on a complete graph of $n$ nodes, where each node (agent/spin/particle) has a discrete state $s_i\in\{1,\ldots,q\}$ representing community membership. The collective behavior is governed by a Hamiltonian $H[\{s\}]$, which assigns an energy to each configuration $\{s\}$. The Hamiltonian for up to $r$-body interactions reads:
\begin{equation}
    H[\{s\}]
    = -\sum_{k=2}^r \frac{J_k}{n^{k-1}} \sum_{i_1<\dots<i_k} \delta(s_{i_1},\ldots,s_{i_k}),
    \label{eq:hamiltonian-r-body}
\end{equation}
where $\delta(s_{i_1},\ldots,s_{i_k}) = 1$ if all arguments are equal, and $0$ otherwise. For clarity, $J_k < 0$ suppresses alignment (repulsion) among $k$-cliques, while $J_k > 0$ promotes alignment (attraction). 

The scaling $1/n^{k-1}$ ensures extensivity: the total energy scales linearly with the system size $n$ as $n \to \infty$. When $r=2$, the system reduces to the conventional $q$-state Potts model with only pairwise interactions. At $r=2$ and $q=2$, we recover the conventional mean-field Ising model with all-to-all couplings.

At equilibrium at temperature $T$ (in energy units), the configurations follow the Boltzmann distribution:
\begin{equation}
    P[\{s\}] = \frac{e^{-H[\{s\}]/T}}{Z_n}, \quad Z_n = \sum_{\{s\}} e^{-H[\{s\}]/T},
    \label{eq:boltzmann-and-Zn}
\end{equation}
where the partition function $Z_n$ sums over all $q^n$ possible spin configurations.

Since the $\delta$-functions in \cref{eq:hamiltonian-r-body} count same-spin cliques, the energy on a complete graph depends only on the occupation numbers $n_1,\ldots,n_q$ ($n=n_1+\cdots+n_q$):
\begin{equation}
    H[\{s\}] = H(n_1,\ldots,n_q) 
    = -\sum_{k=2}^r \frac{J_k}{n^{k-1}}\sum_{p=1}^q \binom{n_p}{k},
    \label{eq:hamiltonian-occup}
\end{equation}
where $\binom{n_p}{k} = n_p(n_p-1)\ldots(n_p-k+1)/k!$ is the binomial coefficient.

For $n\to\infty$, set concentrations $c_p=n_p/n$ and expand \cref{eq:hamiltonian-occup} in $1/n$:
\begin{equation}
    \begin{aligned}
        H_n(c_1, ..., c_q) &= H(nc_1, ..., nc_q) \\
        &= -n \sum_{k=2}^r \sum_{p=1}^q \frac{J_k}{k!} c_p^k + \const + O(1/n),
    \end{aligned}
    \label{eq:hamiltonian-concentration-limit}
\end{equation}
where the $\const$ shift can be omitted and the $O(1/n)$ terms vanish in the thermodynamic limit.

\subsection{Effective potential in terms of occupation numbers}
\label{sec:model-potential-occup}

Summing over microstates with fixed occupation numbers gives the partition function:
\begin{equation}
    Z_n = \sum_{n_1+\cdots+n_q=n} \binom{n}{n_1,...,n_q}\ e^{-H(n_1, ..., n_q)/T},
    \label{eq:partition-function-H}
\end{equation}
with $\binom{n}{n_1,...,n_q} = n!/(n_1!\cdots n_q!)$. We define the per-site potential $F(n_1,\ldots,n_q)$ by
\begin{equation}
    Z_n = \sum_{n_1+\cdots+n_q=n} e^{-n\, F(n_1, ..., n_q)/T}.
    \label{eq:partition-function-F}
\end{equation}
Matching \cref{eq:partition-function-H,eq:partition-function-F} term-by-term gives
\begin{multline}
    F(n_1,...,n_q) = \frac{H(n_1, ..., n_q)}{n}
    - T\frac{\ln \Gamma(1+n)}{n} \\
    + \sum_{p=1}^q T\frac{\ln \Gamma(1 + n_p)}{n},
    \label{eq:F-finite-n}
\end{multline}
($\ln\Gamma$ is the log-Gamma function). Writing $F_n(c_1,\ldots,c_q)\equiv F(nc_1,\ldots,nc_q)$ and using Stirling's formula, together with \cref{eq:hamiltonian-concentration-limit}, gives
\begin{equation}
    \begin{aligned}
        F_n(c_1,...,c_q)
        &= -\sum_{p=1}^q \sum_{k=2}^r \frac{J_k}{k!} c_p^k
        + T \sum_{p=1}^q c_p \log c_p \\
        &\quad + O\left(\frac{\log n}{n}\right).
    \end{aligned}
    \label{eq:potential-concentration-limit}
\end{equation}

\cref{eq:potential-concentration-limit} admits a direct thermodynamic interpretation in the limit $n \to \infty$. Let us identify the per-site internal energy:
\begin{equation}
    U_{\infty}(c_1,...,c_q) = -
    \sum_{p=1}^q
    \sum_{k=2}^r
    \frac{J_k}{k!}c_p^k, \quad c_1 + ... + c_q = 1,
    \label{eq:U-infty}
\end{equation}
and the per-site entropy:
\begin{equation*}
    S_{\infty}(c_1,...,c_q) = -\sum_{p=1}^q c_p \log c_p, \quad c_1 + ... + c_q = 1,
\end{equation*}
The potential $F_n$ takes the standard form of the per-site free energy when $n \to \infty$:
\begin{equation}
    \begin{aligned}
        F_{\infty}(c_1,...,c_q) &= -\sum_{p=1}^q \sum_{k=2}^r \frac{J_k}{k!} c_p^k + T \sum_{p=1}^q c_p \log c_p \\
        &= U_{\infty}(c_1,...,c_q) - T S_{\infty}(c_1,...,c_q).
    \end{aligned}
    \label{eq:F-infty}
\end{equation}
For clarity, $F_n(\mathbf{c})$ is the finite-$n$ per-site potential \cref{eq:potential-concentration-limit}; $F_\infty(\mathbf{c})=\lim_{n\to\infty}F_n(\mathbf{c})$ is \cref{eq:F-infty}.

If we add a magnetic field $(h_1,...,h_q)$ acting separately on each spin type, the initial Hamiltonian \cref{eq:hamiltonian-r-body} and the Hamiltonian in terms of occupation numbers from \cref{eq:hamiltonian-occup} acquire additional terms
\begin{equation*}
    -\sum_{p=1}^q h_p \sum_{i=1}^n \delta(s_i, p) = -\sum_{p=1}^q h_p n_p = -n \sum_{p=1}^q h_p c_p.
\end{equation*}
Consequently, the exact Hamiltonian from \cref{eq:hamiltonian-occup} and the asymptotic potential from \cref{eq:F-infty} are equipped with $-\sum_{p=1}^q h_p c_p$. The particular case of the resulting system when the interaction is limited to the two-body coupling ($r=2$) is studied in \cite{lorenzoniExactAnalysisPhase2019a}.

For $n \to \infty$, the large-deviation principle for occupation fractions \cite{touchetteLargeDeviationApproach2009a} implies that $Z_n$ from \cref{eq:partition-function-F} is dominated by occupation fractions minimizing $F_\infty$ on the simplex $\sum_p c_p=1$. In the following sections we therefore analyze global minima (and, where noted, degeneracies) of $F_\infty$ and its reduced form $f_\infty$ introduced below.

\section{Phase Structure in the Thermodynamic Limit}
\label{sec:phase-structure}
\subsection{Two-color system and the pure entropic regime}
\label{sec:q2-critical}

\begin{figure*}[!ht]
    \includegraphics[width=\linewidth]{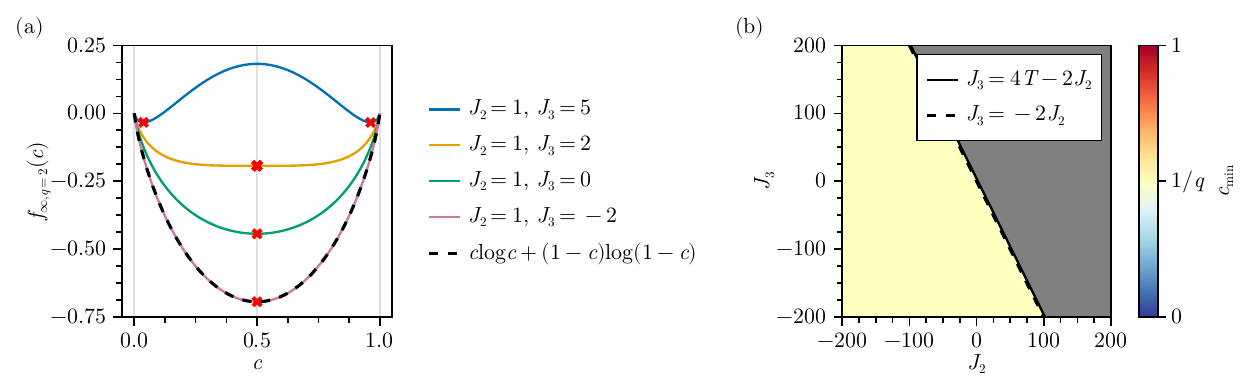}
    \caption{
        \label{fig:q2-phase-and-potential}
        $q=2$, two- and three-body couplings, $T=1$.
        (a) $f_{\infty,q=2}$ from \cref{eq:f-infty-q2} (shifted to $f_{\infty,q=2}(0)=0$); dashed: $c\log c+(1-c)\log(1-c)$; crosses: global minima.
        (b) $(J_2,J_3)$ phase diagram; grey: degenerate minima with $|f_{\infty,q=2}(c^{(1)}_{\min})-f_{\infty,q=2}(c^{(2)}_{\min})|<\degeneps$.
        Solid: $J_3=4T-2J_2$ (phase boundary); dashed: $J_3=-2J_2$ (\cref{eq:pure-entropy-q2-J3=-2J2}).
    }
\end{figure*}

Begin with a $q=2$--color system of $r$-body interacting Ising spins. From \cref{eq:F-infty} the effective potential reads: 
\begin{multline}
    f_{\infty, q=2}(c) \equiv F_{\infty}(c, 1 - c) = 
    -\sum_{k=2}^r \frac{J_k}{k!} \left[(1-c)^k + c^k\right] \\
    + T \left[c \log c + (1-c) \log(1-c)\right].
    \label{eq:f-infty-q2}
\end{multline}
Expression \cref{eq:f-infty-q2} is invariant under $c \leftrightarrow 1-c$, so $c=1/2$ is a symmetric stationary point with $f'_{\infty,q=2}(1/2)=0$ (see \cref{fig:q2-phase-and-potential}(a)). Since $f'_{\infty,q=2}|_{c=0^+}=-\infty$ and $f'_{\infty,q=2}|_{c=1^-}=+\infty$, the global minimum is not attained at the boundary for $T>0$, although it can be approached arbitrarily closely as $T\to 0^+$. If $f''_{\infty,q=2}(1/2)<0$, the symmetric point is a local maximum. The global minimum is then attained at (at least) two points symmetric around $c=1/2$, and the system is degenerate. For $r=3$, the condition $f''_{\infty,q=2}(1/2)=0$ defines the phase boundary $J_3 = 4T - 2J_2$ (solid line in \cref{fig:q2-phase-and-potential}(b)), where the symmetric state loses stability and a pair of minima $\{c_{\min},1-c_{\min}\}$ emerges continuously.

Couplings $J_k$ of odd orders $k$ do not contribute to $c^k$ in \cref{eq:f-infty-q2}.  Thus, for odd $r=2m+1$, the energy polynomial is a polynomial of order $r-1$:
\begin{multline*}
    U_{\infty}(c, 1-c)\big|_{r \equiv 1 \hspace{-0.2cm}\pmod{2}} = \ldots \\
    + \frac{2J_{r-1} + J_r}{2(r-2)!} c^{r-2} - \frac{2J_{r-1} + J_r}{(r-1)!}c^{r-1}.
\end{multline*}
The polynomial has $r-2$ relevant coefficients (again, a constant term in energy does not affect statistical properties of the Boltzmann distribution). Thus, one can find a $(\geq 1)$-dimensional subspace in the $r-1$-dimensional phase space of couplings $J_2,...,J_r$ which makes $U_{\infty}(c, 1-c)$ independent of $c$. 
For example, in case of only two- and three-body interactions ($r=3$) one has (see \cref{fig:q2-phase-and-potential})
\begin{multline}
    F_{\infty}(c, 1-c)\big|_{r=3} = \const + T \left[c \log c + (1-c) \log(1-c)\right], \\
    \text{when}\ J_3=-2J_2, J_4=J_5=\cdots=0.
    \label{eq:pure-entropy-q2-J3=-2J2}
\end{multline}
Proceeding recursively such coupling subspace can be built for any odd $r$. In other words, there is a coupling submanifold on which the potential $f_{\infty,q=2}(c)=F_{\infty}(c, 1 - c)$ coincides with the Potts mixing entropy alone (up to a constant shift), while the microscopic Hamiltonian \cref{eq:hamiltonian-r-body} need not be constant in individual configurations.

\subsection{General $q$-color system: Symmetric ansatz and connection to the $p$-star model}
\label{sec:general-q-and-p-star}

Since $F_\infty$ is $S_q$-invariant, we search for minimizers on the symmetric manifold
\begin{equation*}
    c_1=c,\ c_2=\ldots=c_q=\frac{1-c}{q-1}, \quad c \in [0, 1].
\end{equation*}
The optimization thus reduces to studying the univariate function:
\begin{equation}
    f_n(c) = F_n\left(c, \frac{1-c}{q-1}, \ldots, \frac{1-c}{q-1}\right)
    \label{eq:reduced-potential-fn}
\end{equation}
where the permutation symmetry ensures that this reduction is independent of the specific color chosen to have $c_1 = c$. For $r=2$, global minimizers of $F_\infty$ on the simplex lie on this ansatz \cite{wuPottsModel1982,georgiiMeanfieldTheoryPotts2006,lorenzoniExactAnalysisPhase2019a}.
For $r\ge 3$, we adopt the ansatz throughout \cref{sec:phase-structure} and check in \cref{fig:configurations-and-landscapes} (lower panels) that full simplex minimization of $F_\infty$ agrees with $f_\infty$.

In the thermodynamic limit we can substitute \cref{eq:F-infty} into \cref{eq:reduced-potential-fn} and split the resulting expression into two parts:
\begin{multline}
    f_{\infty}(c) = \underbrace{- \sum_{k=2}^r \frac{J_k}{k!} \left[\frac{(1-c)^k}{(q-1)^{k-1}} + c^k\right] + c T \log(q-1)}_{\text{polynomial in $c$}} \\
    + T \underbrace{\left[c \log c + (1-c) \log(1-c)\right]}_{-\text{entropy in $c$}}, \quad c \in [0, 1],
    \label{eq:reduced-potential-finf}
\end{multline}
where we omitted the irrelevant constant shift. The linear term $c T \log(q-1)$ comes from the $q$-degeneracy of the system and vanishes when $q=2$. The highest order term in \cref{eq:reduced-potential-finf} is 
\begin{equation*}
    c^r \frac{J_r}{r!} \left(\frac{(-1)^r}{(q-1)^{r-1}} + 1\right).
\end{equation*}
For $q \neq 2$ this term cancels out only if $J_r = 0$. Thus to cancel out the $c^{r-1}$-term, which depends on $J_r$ and $J_{r-1}$, we need to set $J_{r-1} = 0$. Proceeding recursively we see that eliminating the $c$-dependence of the energy in \cref{eq:reduced-potential-finf} for $q \neq 2$ requires zero coupling constants. 
Thus the system with non-trivial interactions may become purely entropic only in case of $q=2$ colors (see the discussion around \cref{eq:pure-entropy-q2-J3=-2J2}).

\begin{figure}[!htb]
    \includegraphics[width=\linewidth]{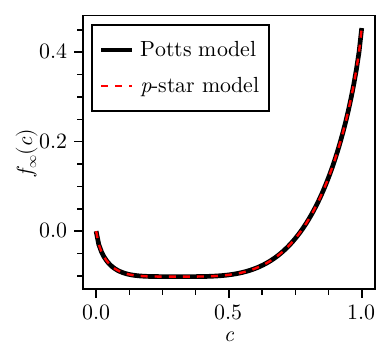}
    \caption{
        \label{fig:poly-param-vs-p-star}
        Match of $f_\infty$ (\cref{eq:reduced-potential-finf}; $q=3.05019$, $J_2=6.93398$, $J_3=-11.9041$) and $p$-star form (\cref{eq:reduced-potential-poly}; $t_i$ from \cite{biondiniPstarModelsMeanfield2022}) at $T=1$, via \cref{eq:Potts-pstar-mapping}.
    }
\end{figure}

Now consider \cref{eq:reduced-potential-finf}, treating $q$ as a continuous degree of freedom ($T$ is fixed). 
Together with $J_2,...,J_r$ we have $r$ free parameters that form a polynomial in $c$ of order $r$:
\begin{equation}
    f_{\infty}(c)\big|_{q\in\R} = -2\sum_{k=1}^r t_k c^k + T \left[c \log c + (1-c) \log(1-c)\right],
    \label{eq:reduced-potential-poly}
\end{equation}
where $t_1,...,t_r$ depend on $q,J_2,...,J_r$. Reverting logic, we can find specific values of $q,J_2,...,J_r$ for a given set of constants $t_1,...,t_r$. Indeed, if we compare term-by-term \cref{eq:reduced-potential-finf} and \cref{eq:reduced-potential-poly}, we obtain a system of equations. For simplicity, consider case $r=3$:
\begin{equation}
    \begin{dcases}
        2t_1 = -T\log(q-1) - \frac{J_2}{q-1} - \frac{J_3}{2(q-1)^2} \\
        2t_2 = \frac{J_2}{2(q-1)} + \frac{J_2}{2} + \frac{J_3}{2(q-1)^2} \\
        2t_3 = -\frac{J_3}{6(q-1)^2} + \frac{J_3}{6}
    \end{dcases},
    \label{eq:Potts-pstar-mapping}
\end{equation}
from which follows an implicit equation for $q$:
\begin{equation}
    2t_1 = -T \log(q-1) - \frac{4t_2}{q} - \frac{6t_3(1-2/q)}{(q-1)^2 - 1}.
    \label{eq:Potts-pstar-implicit-q}
\end{equation}
Analogous implicit equations can be derived for any $r > 3$. We solve \cref{eq:Potts-pstar-implicit-q} numerically by minimizing the squared residual with a multistart bounded search. Once solved for $q$, the couplings $J_2$ and $J_3$ are found directly from \cref{eq:Potts-pstar-mapping} (see \cref{fig:poly-param-vs-p-star}). Due to the log-term, matching \cref{eq:Potts-pstar-mapping} for general $(t_1,\ldots,t_r)$ often requires non-integer $q$: a formal continuation of the reduced potential, not a physical number of communities.

Here we connect the reduced potential \cref{eq:reduced-potential-poly} with a mean-field model of the $p$-star exponential random graph \cite{biondiniPstarModelsMeanfield2022} (in our notation $p=r$). That is, if we replace the concentration $c$ by the mean connectance (mean fraction of the present edges) defined as $\mean{L} = \sum_{ij} A_{ij}/(n(n-1)) \in[0, 1]$ of a random Boolean adjacency matrix $(A_{ij})$, we restore the same expression as in \cite{biondiniPstarModelsMeanfield2022} for the free energy in a random graph model, where all $k \leq r$ tuples of edges interact with the coupling constant $t_k$. Thus the mapping is exact for the reduced potentials \cref{eq:reduced-potential-finf,eq:reduced-potential-poly}, but not for the microscopic Potts Hamiltonian \cref{eq:hamiltonian-r-body} and does not admit a direct parameter-wise interpretation.
For a more detailed analysis of \cref{eq:reduced-potential-poly} in the case of $r=3$, see \cite{biondiniPstarModelsMeanfield2022}.

\subsection{System with two- and three-body interactions}
\label{sec:two-and-three-body}

\begin{figure*}[!htb]
    \includegraphics[width=\linewidth]{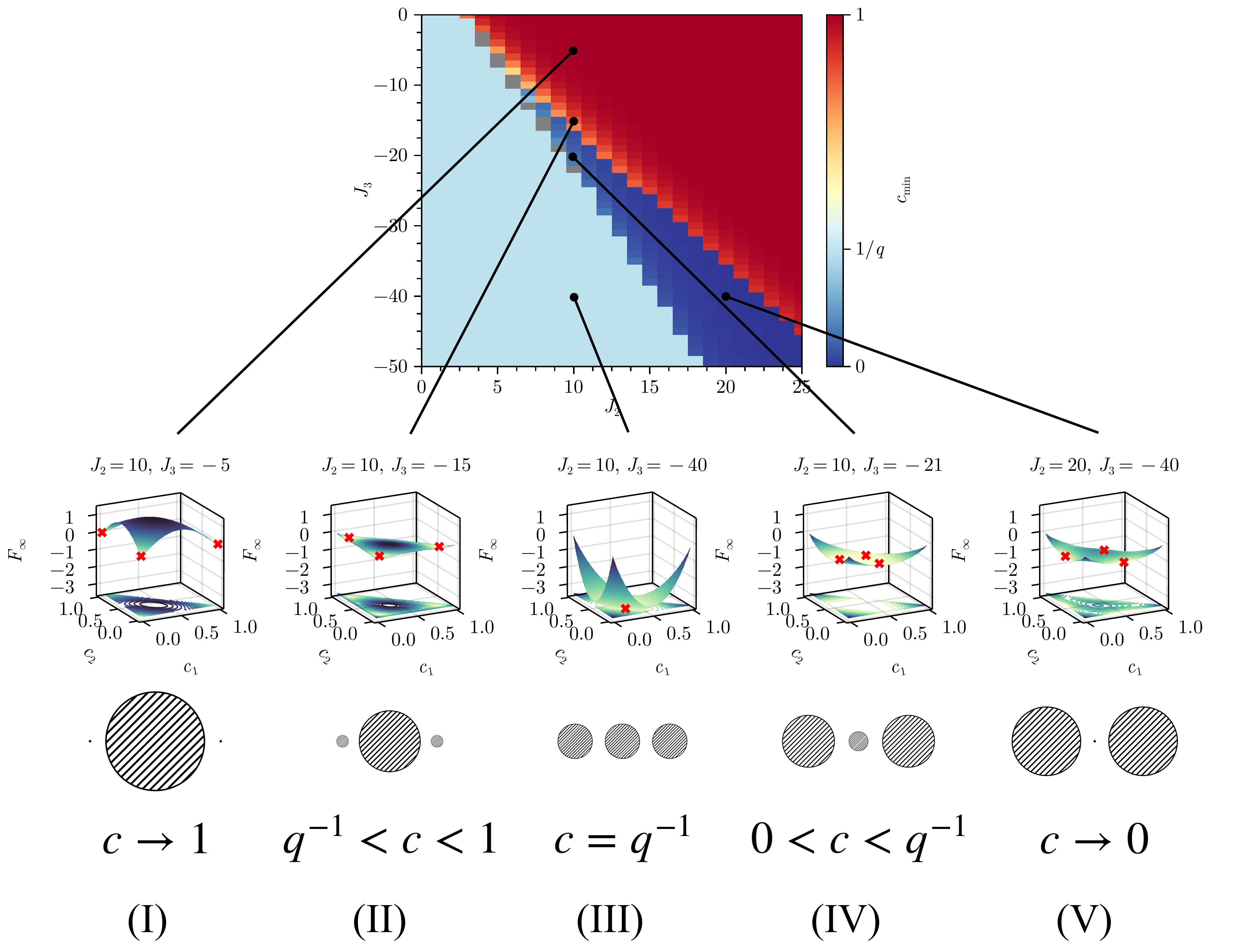}
    \caption{
        \label{fig:configurations-and-landscapes}
        $q=3$, $T=1$: phase diagram from $f_\infty$ in \cref{eq:reduced-potential-finf}.
        $c_{\min}$ is the distinguished color fraction at the global minimum of $f_\infty$.
        Grey: degenerate global minima ($|f_{\infty}(c^{(1)}_{\min})-f_{\infty}(c^{(2)}_{\min})|<\degeneps$).
        Lower panels: $F_\infty$ on the simplex (shifted: $F_\infty(1,0,0)=0$); crosses: minima of $f_\infty(c)=F_\infty(c,\tfrac{1-c}{2},\tfrac{1-c}{2})$.
        Regimes (I)--(V) as in \cref{sec:two-and-three-body}.
    }
\end{figure*}

The global minima of the reduced potential $f_\infty(c)$ (defined in \cref{eq:reduced-potential-finf}) depend critically on the interaction coefficients $J_2, \dots, J_r$ and fixed $T$. Taking our illustrative case with two- and three-body interactions, we plot the phase diagram in the $(J_2, J_3)$ plane in \cref{fig:configurations-and-landscapes} for $q=3$ colors. The minimizers of $f_\infty$ realize five regimes (sociological readings in \cref{sec:discussion-sociology}):
\begin{enumerate}[(I)]
    \item \textbf{Complete-consensus} ($c_{\min}\to 1$): one color carries almost all weight.
    \item \textbf{Consensus} ($1/q < c_{\min} < 1$): one majority color with equal minorities at concentration~\mbox{$(1-c_{\min})/(q-1)$}.
    \item \textbf{Symmetric} ($c_{\min} = 1/q$): all $q$ colors equally populated.
    \item \textbf{Reduced-symmetry} ($0 < c_{\min} < 1/q$): one depleted color and $q-1$ equal others; this interior branch is absent at $T=0$.
    \item \textbf{Complete reduced-symmetry} ($c_{\min}\to 0$): only $q-1$ colors survive with equal weight.
\end{enumerate}
Types are: consensus ($c_{\min}>1/q$), symmetric ($c_{\min}=1/q$), and reduced-symmetry ($c_{\min}<1/q$). The energy term in \cref{eq:reduced-potential-finf} is regular in $c$, so the entropy piece dominates $df_\infty/dc$ near $c=0$ and $c=1$, so regimes (I) and (V) are not realized as $c_{\min}\in\{0,1\}$ at $T>0$. Yet, this difference can be negligible in practice (for example, at the scale of \cref{fig:configurations-and-landscapes}).
Moreover, finite-size effects differ between scenarios (I) and (II) (and also between (IV) and (V)), so we treat them separately.

\begin{figure*}[!ht]
    \includegraphics[width=\linewidth]{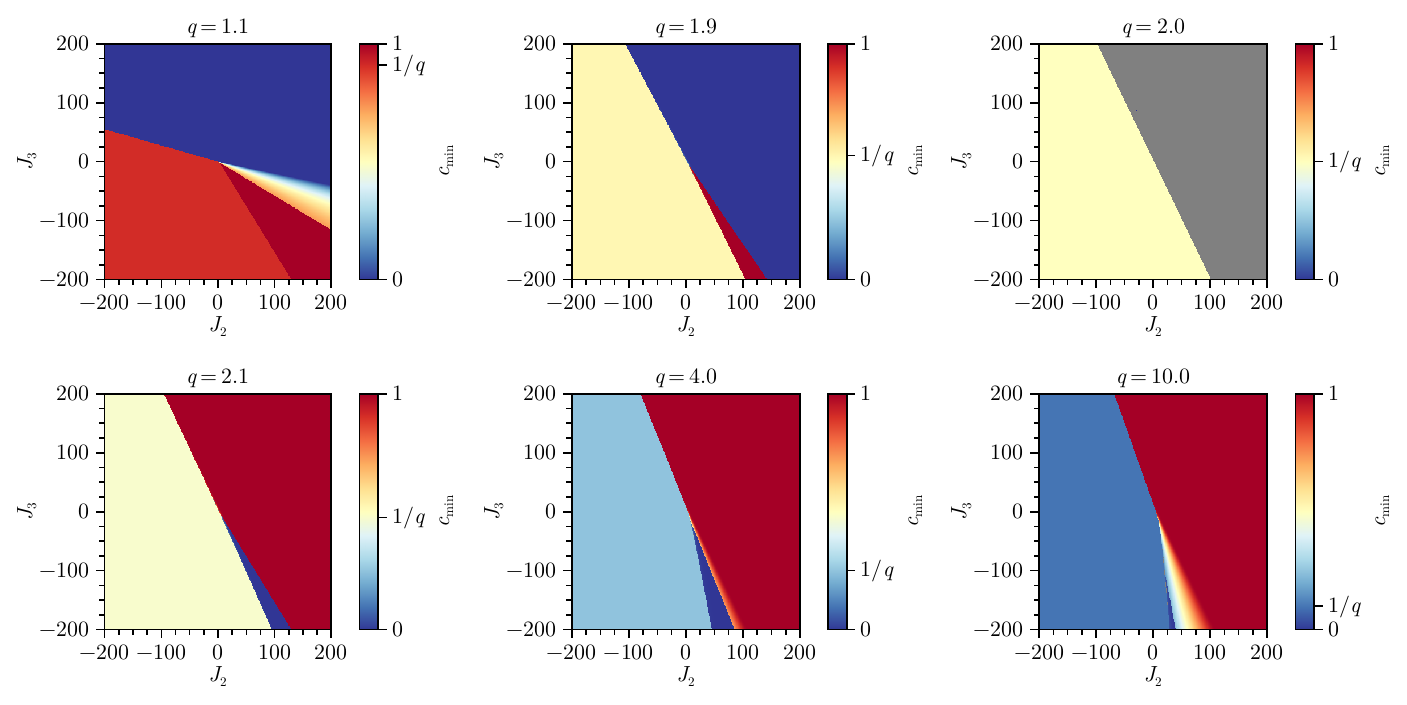}
    \caption{
        \label{fig:phase-diagrams-vs-q}
        $T=1$ phase diagrams from $f_\infty$ (\cref{eq:reduced-potential-finf}) for several $q$.
        Grey: degenerate global minima ($|f_{\infty}(c^{(1)}_{\min})-f_{\infty}(c^{(2)}_{\min})|<\degeneps$).
    }
\end{figure*}

In \cref{fig:phase-diagrams-vs-q} we show phase diagrams for different values of $q$. A richer phase structure appears in the quadrant with $J_2 > 0$ (pair attraction) and $J_3 < 0$ (triadic repulsion), where these interactions compete strongly.

\section{Finite-Size Effects and Monte-Carlo Simulations}
\label{sec:finite-effects}

\begin{figure*}[!ht]
    \includegraphics[width=\linewidth]{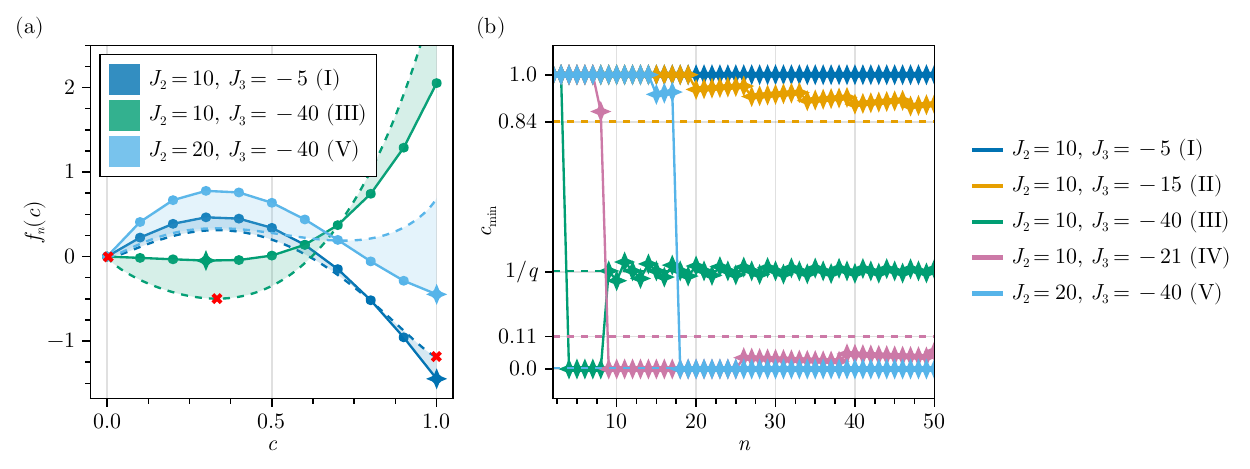}
    \caption{
        \label{fig:discrete-effects}
        $q=3$, $T=1$: finite-$n$ commensurability (phases classified in \cref{fig:configurations-and-landscapes}).
        (a) $f_n$ on the grid $c=i/n$ (solid, discrete) and $f_\infty$ (dashed) vs.\ $c$, both shifted to $0$ at $c=0$; crosses: $f_\infty$ minima; stars: $f_n$ minima.
        (b) $c_{\min}(n)$ from $f_n$; dashed: $f_\infty$ limits.
    }
\end{figure*}

For finite $n$, only discrete concentrations $c_p\in\{0,1/n,\ldots,1\}$ are realizable, so minimizers of $F_n$ on the grid need not coincide with minimizers of $F_\infty$ (so called commensurability effects). 
\cref{fig:discrete-effects}a shows these effects for $q=3$ at $T=1$ (stars vs.\ crosses). For $J_2=20$, $J_3=-40$, a thermodynamic-limit point in regime (V) can correspond at finite $n$ to a minimum associated with another regime (light-blue region in \cref{fig:discrete-effects}(a)). Different regimes also respond differently to $n$ (\cref{fig:discrete-effects}(b)).

\begin{figure}[!ht]
    \includegraphics[width=\linewidth]{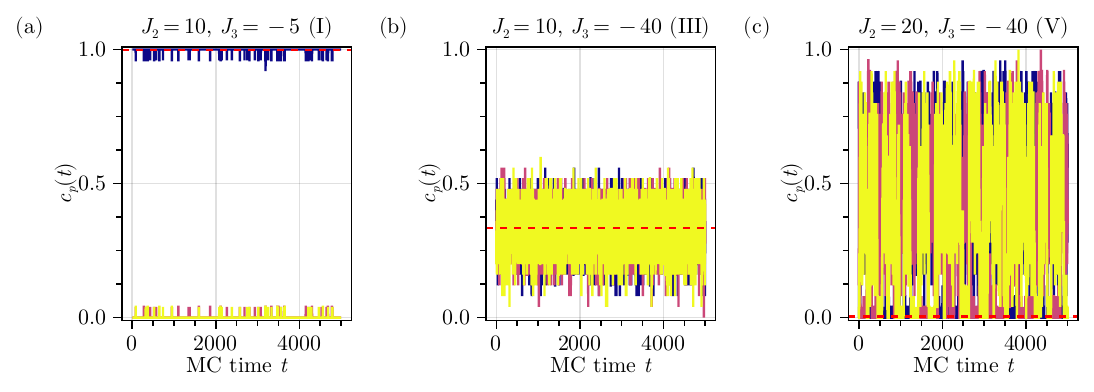}
    \caption{
        \label{fig:mc-simulations}
        MC: $n=25$, $5000$ sweeps, $T=1$ (see text); dashed: $f_\infty$ minima ($<\degeneps$ depth tolerance).
        (a) $J_2=10$, $J_3=-5$, regime (I); (b) $J_2=10$, $J_3=-40$, regime (III); (c) $J_2=20$, $J_3=-40$, regime (V).
    }
\end{figure}

\begin{figure}[!ht]
    \includegraphics[width=\linewidth]{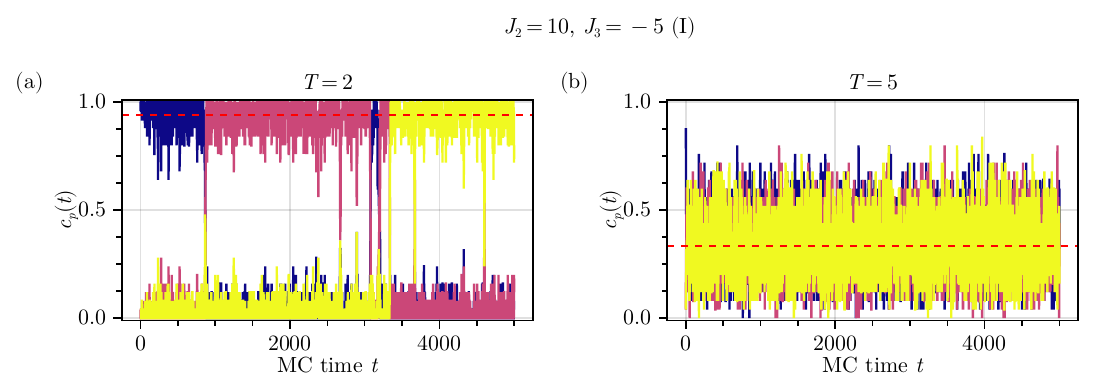}
    \caption{
        \label{fig:mc-temperature-jumps}
        MC with $J_2=10$, $J_3=-5$: (a) $T=2$; (b) $T=5$; dashed: $f_\infty$ minima at each $T$.
    }
\end{figure}

We perform Monte Carlo (MC) simulations of the complete-graph $q=3$ Potts model at fixed $J_2$, $J_3$ using single-spin Metropolis updates on the occupation-number energy \cref{eq:hamiltonian-occup}: each sweep proposes one random recoloring per site (accepted with probability $\min\{1,e^{-\Delta H/T}\}$). Trajectories use $n=25$ spins, initial condition $s_i=1$ for all $i$, $5000$ sweeps, and record $c_p(t)$ after every sweep. Figures~\ref{fig:mc-simulations}--\ref{fig:mc-temperature-jumps} show one representative trajectory per panel.

MC samples the Boltzmann ensemble \cref{eq:boltzmann-and-Zn}; plotted trajectories $c_p(t)$ are instantaneous fractions after each sweep, not time averages of $F_n$, nor deterministic minimizers of $F_\infty$.
However, the thermodynamic-limit analysis is still valid in the leading order, and we can speak of phase-like regimes rather than strict thermodynamic phases.

From \cref{fig:discrete-effects}, $n=25$ is large enough that the minimizers of $f_n$ close to those of $f_\infty$ for $J_2=10$, $J_3=-5$ (regime I), $J_2=10$, $J_3=-40$ (III), and $J_2=20$, $J_3=-40$ (V). In regime (I) (\cref{fig:mc-simulations}(a), $T=1$), $c_p(t)$ stays near the complete-consensus minimum and the dominant color does not change: barriers between the three equivalent consensus wells are not overcome on the simulation timescale (broken ergodicity). In regime (III) (\cref{fig:mc-simulations}(b)), the trajectory fluctuates around $c_p\approx 1/3$. In regime (V) (\cref{fig:mc-simulations}(c)), the landscape is shallower (\cref{fig:configurations-and-landscapes}) and the trajectory hops between degenerate reduced-symmetry states (note a color at each time step that has fraction close to $0$).

Raising $T$ increases both thermal barrier crossing and the reshaping of $f_\infty$ towards the entropy-dominated regime (\cref{eq:reduced-potential-finf}). In \cref{fig:mc-temperature-jumps}(a), where $T=2$, the dominant color switches more frequently during the simulation window than in \cref{fig:mc-simulations}(a), where $T=1$.
A system in the complete-consensus regime (I) at $T=1$ moves to the consensus regime (II) at $T=2$ (red dashed line in \cref{fig:mc-temperature-jumps}(a)) and reaches the symmetric regime (III) at $T=5$ (see \cref{fig:mc-temperature-jumps}(b)).

\section{Social--Quantum Correspondence}
\label{sec:social-quantum-corr}

Having established the phase structure of the Potts model (\cref{sec:phase-structure}), we now demonstrate a mean-field correspondence with the $SU(N)$ spin systems \cite{polychronakosFerromagneticPhaseTransitions2023, polychronakosPhaseTransitionsDecomposition2024, polychronakosNonabelianFerromagnetsThreebody2025}. For completeness, we briefly recap some parts from those references that are essential for our mapping.

The $SU(N)$-symmetric Hamiltonian describes $n$ atoms on a lattice with two- and three-body internal couplings.
At site $i$, generators $S_i^a$ ($a=1,\ldots,N^2-1$) satisfy $[S_i^a,S_j^b]=i\delta_{ij}\sum_c f^{abc}S_i^c$, where $f^{abc}$ are the structure constants of $SU(N)$. The Hamiltonian is
\begin{equation}
    \mathcal{H} = -\sum_{i<j} \sum_{a=1}^{N^2-1} J_{ij} S_i^a S_j^a - \sum_{i<j<k} \sum_{a,b,c=1}^{N^2-1} J_{ijk} d_{abc} S_i^a S_j^b S_k^c,
    \label{eq:suN-spin-J2-J3}
\end{equation}
with a symmetric tensor $d_{abc}$ and fully symmetric couplings $J_{ij},J_{ijk}$. In the mean-field reduction used in \cite{polychronakosNonabelianFerromagnetsThreebody2025},
\begin{equation}
    \mathcal{H} = -\frac{\alpha}{n}C^{(2)} -\frac{\beta}{n^2}C^{(3)},
    \label{eq:interactions}
\end{equation}
where $C^{(2)},C^{(3)}$ are the Casimir operators of $SU(N)$ and $\alpha,\beta$ are the lattice-dependent coefficients.

Irreducible representations of $SU(N)$ are labeled by integer partition data $k_i$ ($i=1,\ldots,N$) with $\sum_i k_i=n+N(N-1)/2$, related to Young-tableau row lengths $l_i$ \cite{polychronakosNonabelianFerromagnetsThreebody2025, knappLieGroupsIntroduction2005}. In that parametrization,
\begin{equation*}
    C^{(2)} = \frac{1}{2}\sum_{i=1}^N k_i^2 + \text{const}, \quad C^{(3)} = \frac{1}{6}\sum_{i=1}^N k_i^3 + \text{const}.
\end{equation*}
The partition function is $Z=\sum_\lambda \dim(V_\lambda)\dim(\mathcal{M}_\lambda)e^{-E_\lambda/T}$, with
\begin{equation*}
    \begin{aligned}
        \dim(V_\lambda) &= \prod_{1 \leq i<j\leq N} \frac{l_i - l_j + j - i}{j - i}, \\
        \dim(\mathcal{M}_\lambda) &= \frac{n!}{\prod_i k_i!} \prod_{i<j} (k_i - k_j)
    \end{aligned}
\end{equation*}
(see \cite{polychronakosPhaseTransitionsDecomposition2024, polychronakosNonabelianFerromagnetsThreebody2025}).
Stirling's formula applied to $\dim(\mathcal{M}_\lambda)$ reveals the same multinomial entropy structure as in \cref{eq:partition-function-H,eq:F-infty}. In the thermodynamic limit the mean-field free energy per site is
\begin{equation}
    \mathcal{F} = - \frac{\alpha}{2}\sum_{i=1}^N x_i^2 - \frac{\beta}{6}\sum_{i=1}^N x_i^3 
    + T\sum_{i=1}^N x_i \ln x_i,
    \label{eq:SU(N)}
\end{equation}
where $\sum_i x_i=1$ and the $x_i=k_i/n$ are order parameters (normalized shifted Young diagram row lengths).

Truncating the Potts model at $r=3$ and setting $q=N$, term-by-term comparison of \cref{eq:SU(N)} with \cref{eq:F-infty} yields the coupling identification
\begin{equation}
    \alpha = J_2, \qquad \beta = J_3,
    \label{eq:potts-su-couplings}
\end{equation}
with $c_p \leftrightarrow x_p$.

Both systems show the same pattern of $S_q\simeq S_N$ symmetry breaking between a high-temperature symmetric sector ($c_p=1/q$ / $x_i=1/N$) and low-temperature sectors with one depleted mode. The work
\cite{polychronakosNonabelianFerromagnetsThreebody2025} describes this as an $SU(N)\to SU(N-1)\times U(1)$ degeneration; regimes (IV)--(V) in \cref{fig:configurations-and-landscapes} are the analogous Potts configurations with one depleted color.

\section{Discussion}
\label{sec:discussion}

\subsection{Sociological interpretation of phase transitions in the $q$-state Potts model on a complete graph}
\label{sec:discussion-sociology}

We analyze a $q$-state Potts model with many-body interactions on a complete graph, motivated by the work \cite{avetisovPhaseTransitionsSocial2018a} on Schelling-type dynamics on networks. The complete graph topology has two roles: mathematically, it makes mean-field arguments exact; sociologically, it approximates highly connected (\enquote{transparent}) societies. Our analysis reveals collective equilibrium-clustering patterns in this regime of maximal connectivity and information exposure.

As shown in \cref{sec:two-and-three-body,fig:configurations-and-landscapes}, the model exhibits three principal equilibrium types (with two entropy-limited variants): symmetric, reduced-symmetry and consensus. Phase boundaries in \cref{fig:configurations-and-landscapes} can be read as sociopolitical shifts associated with spontaneous breaking of the $S_q$ permutation symmetry by equilibrium configurations.

In the complete-consensus regime (I), almost all weight sits on one color. In the consensus regime (II), one state is clearly dominant with equal minorities. In social terms, (I)--(II) describe societies where one ideology or cultural identity becomes hegemonic, with minorities present but comparatively weak.

In the symmetric regime (III), all $q$ communities are equally represented. It corresponds to a \enquote{democratic} configuration where social power is distributed equally across $q$ distinct communities (or viewpoints). 

In the reduced-symmetry regime (IV) and the complete reduced-symmetry limit (V), one state is depleted and $q-1$ communities remain equally populated. Sociologically, this represents a society where one group becomes culturally disengaged, marginalized, or assimilated. Though the structure appears pluralistic, underlying asymmetry hints at latent tensions or loss of voice for one population segment. The correspondence with the $SU(N)$ quantum spin system (\cref{sec:social-quantum-corr}) describes the same order-parameter pattern as $SU(N) \to SU(N-1) \times U(1)$ in \cite{polychronakosNonabelianFerromagnetsThreebody2025}. A related manifestation of symmetry breaking on complete (\enquote{transparent}) graphs is segregation in Schelling-type dynamics~\cite{stoicaSchellingSchoolsComparison2014}.

\subsection{The entropy-driven regime as an atomized society}
\label{sec:entropy-driven}

\cref{eq:pure-entropy-q2-J3=-2J2} identifies a special regime for $q=2$ when $J_3 = -2J_2$. At this balance, the interaction contribution cancels out, and collective behavior is governed solely by the entropy of mixing.
In physical terms, this corresponds to a purely disordered state with no concentration-dependent energetic bias. 
Sociologically, this suggests a limiting regime in which structured social influence is weak and individual opinion is driven largely by randomness.

This regime closely resembles an \enquote{atomized} society, a concept rooted in classical sociology. In such societies, individuals are no longer integrated into strong collective identities or institutions. The weakening of social connections --- typically maintained by religion, stratification, or tradition --- leads to what Durkheim termed \enquote{anomie} \cite{durkheimSuicideStudySociology1951}. This is a state of "normlessness" where individuals behave as isolated agents, forming opinions independently of shared norms. The resulting distribution of opinions has no interaction-driven bias toward consensus.

Some features of contemporary digital environments are qualitatively consistent with this entropy-driven regime.
It echoes, in a speculative sense, \enquote{postmodernity}, where identities are fluid, social ties are optional, and meaning is increasingly self-constructed. In this sense, the entropy-dominated regime in our model captures a sociological insight: the screening of structured interactions. This analogy is interpretive and does not immediately imply a one-to-one causal mapping.

\subsection{Social--quantum correspondence}
\label{sec:corr-outlook}

The expression from \cref{eq:potts-su-couplings} matches the Potts mean-field free energy \cref{eq:F-infty} at $r=3$ to the $SU(N)$ functional \cref{eq:SU(N)}. Within the complete-graph setting, this correspondence links the sociological order parameters $c_p$ to the quantum mean-field variables $x_i$ and to Young-tableau representation. This connection has the following important consequences:
\begin{itemize}
    \item Allows for the classification of phases in the Potts diagram (\cref{fig:configurations-and-landscapes}) in terms of the $SU(N)$ mean-field analysis \cite{polychronakosNonabelianFerromagnetsThreebody2025}.
    \item Provides the representation-theoretic description of Potts macrostates for symmetric and symmetry-broken sectors.
    \item Connects finite-$n$ corrections $O((\log n)/n)$ terms in \cref{eq:potential-concentration-limit} and the quantum multiplicity factor $\dim(\mathcal{M}_\lambda)$.
\end{itemize}

\subsection{Limitations and Future Directions}
\label{sec:limitations-and-future}

While our model is exact on the complete graph within the occupation-number reduction, we identify several limitations. 

First, for $r=3$ the full $F_\infty$ had global minimizers predicted by the symmetric ansatz (see \cref{fig:configurations-and-landscapes}). However, for $r\ge 4$ other symmetric subgroups of $S_q$ may reveal other global minimizers of $F_\infty$ reflecting a more complicated structure of polynomials of higher orders. 

Second, real social networks exhibit complex topologies with clustering, degree heterogeneity, and community structure that deviate significantly from all-to-all connectivity. 

Third, MC at fixed $(n,T)$ does not need to equilibrate among all symmetry-related minima on the timescale simulated (\cref{sec:finite-effects}); real social systems may likewise exhibit history dependence beyond Boltzmann statistics.

Fourth, our model treats agents as distinguishable only by community membership, neglecting individual heterogeneity in influence, or opinion formation mechanisms.
 
The $q$-state Potts model with pairwise interactions is dual to the random-cluster model. An open direction is to investigate whether an analogous representation can be constructed for the present many-body couplings, potentially connecting to metastable switching results \cite{blancaDynamicsMeanfieldRandomcluster2014} and to field-theoretic treatments at non-integer $q$ \cite{delfinoPottsQcolorField2011}.

These limitations define future research directions for sparse networks, full-simplex minimization at $r\ge 3$, non-equilibrium dynamics, and heterogeneous agents, with the social--quantum correspondence as a common analytical framework.

\section*{Data and Code Availability}
\label{sec:data-code-availability}

Source code and data are openly available in \cite{transparentPottsModel2026}.

\begin{acknowledgments}
Inspired by \cite{avetisovPhaseTransitionsSocial2018a}, we thank A. Gorsky, M. Tamm, and O. Valba for fruitful discussions. We express our special gratitude to O. Evnin for critical remarks and important suggestions at different stages of the work. We are grateful to A. Polychronakos for sharing results on many-body quantum spin chains prior to publication, which motivated the social--quantum correspondence in \cref{sec:social-quantum-corr}.
We thank anonymous reviewers whose comments improved the manuscript. Generative AI-based writing tools were only used for stylistic and translational improvements of the manuscript. SN thanks the Beijing Institute for Mathematical Sciences and Applications (BIMSA) for hospitality where part of this work was carried out.
\end{acknowledgments}

\bibliographystyle{apsrev4-2}
\bibliography{biblio}

\end{document}